\documentstyle[prl,aps,preprint,psfig]{revtex}

\begin{document}           
\draft                                          
\title{Spin accumulation induced resistance in mesoscopic ferromagnet/
superconductor junctions}

\author{F.J. Jedema\cite{email}, B.J. van Wees, B.H. Hoving, A.T. Filip, 
and T.M. Klapwijk\cite{teun}}
\address{Department of Applied Physics and Materials Science Center, University 
of Groningen,\\
Nijenborg 4, 9747 AG Groningen, The Netherlands \\} 
\date{\today}

\maketitle

\begin{abstract}
We present a description of spin-polarized transport in mesoscopic
ferromagnet-superconductor (F/S) systems, where the transport is diffusive, and 
the interfaces are transparent. It is shown that the spin reversal associated 
with Andreev reflection generates an excess spin density close to the F/S 
interface, which leads to a spin contact resistance. Expressions for the 
contact resistance are given for two terminal and four terminal geometries. In 
the latter the sign depends on the relative magnetization of the ferromagnetic 
electrodes. 

\end{abstract}


Andreev reflection \cite{Andreev}($AR$) is the elementary process which enables 
electron transport across a normal metal-superconductor (N/S) interface, for 
energies below the superconducting energy gap $\Delta$. The incoming electron 
with spin-up takes another electron with spin-down to enter the superconductor 
as a Cooper pair with zero spin. This corresponds to a reflection of a 
positively charged hole with a reversed spin direction. 

The spin reversal has important consequences for the resistance of a 
ferromagnetic-superconductor (F/S) interface. A suppression of the transmission 
coefficient has been reported in F/S multilayers\cite{aarts1}, and in 
transparent ballistic F/S point contacts a reduction of the conductance has 
been predicted and observed\cite{been1,soulen1,upadhyay1}. In F/S point 
contacts the Andreev reflection process is limited by the lowest number of the 
available spin-up and spin-down conductance channels, which are not equal due 
to a separation of the spin bands in the ferromagnet, caused by the exchange 
interaction. However, in most experiments the dimensions 
of the sample exceed the electron mean free path $l_e$, and therefore the 
electron transport cannot be described ballistically. 

We present a description for spin-polarized transport in diffusive F/S 
systems, in the presence of Andreev reflection for temperatures and energies 
below $\Delta$\cite{foot0}. We will show that the $AR$ process at the F/S 
interface causes a spin 
accumulation close to the interface, due to the different spin-up and spin-down 
conductivities $\sigma_{\uparrow}$ and $\sigma_{\downarrow}$ in the ferromagnet. 

In first approximation we will ignore the effects of phase coherence in the 
ferromagnet, which in the presence of a superconductor can give rise to the 
proximity effect\cite{belzig1,gueron1,giroud1,lead1}. 
The spin-flip length ($\lambda_{sf}^F$) of the electrons in 
the ferromagnet, which is the distance an electron can diffuse before its spin 
direction is randomized, is much larger than the exchange interaction length. 
This means that all coherent correlations in the ferromagnet are expected to be 
lost beyond the exchange length, but the spin of the electron is still 
conserved.   
 
Transport in a diffusive metallic ferromagnet is usually described in terms of 
its spin-dependent conductivities $\sigma_{\uparrow,\downarrow}=e^2N_{\uparrow
,\downarrow}D_{\uparrow,\downarrow}$, where $N_{\uparrow,\downarrow}$ are the 
spin-up and spin-down density of states at the Fermi energy and $D_{\uparrow,
\downarrow}$ the spin-up and spin-down diffusion constants
\cite{joh1,son1,valet,hersh}. In a homogeneous 1D-ferromagnet the current 
carried by both spin directions ($j_{\uparrow,\downarrow}$) is distributed 
according to their conductivities: 
\begin{equation}
j_{\uparrow,\downarrow} ~=~ -(\frac{\sigma_{\uparrow,\downarrow}}{e})\frac
{\partial\mu_{\uparrow,\downarrow}}{\partial x}~  
\label{current}                    
\end{equation} where $\mu_{\uparrow,\downarrow}$ are the electrochemical 
potentials of the spin-up and spin-down electrons, which are equal in a 
homogeneous system. In a non-homogeneous system however, where current is 
injected into, or extracted from a material with different spin-dependent 
conductivities, the electrochemical potentials can be unequal. This is a 
consequence of the finite spin-flip scattering time $\tau_{sf}$, which is 
usually considerably longer than the elastic scattering time $\tau_e$. The 
transport equations therefore have to be supplemented by:
\begin{equation}
D\frac{\partial^2(\mu_\uparrow-\mu_\downarrow)}{\partial^{2} x}~=~\frac
{\mu_\uparrow-\mu_\downarrow}{\tau_{sf}}~ 
\label{diffusion}
\end{equation} where $D=(\frac{N_\downarrow}{(N_\uparrow+N_\downarrow)
D_\uparrow}+ \frac{N_\uparrow}{(N_\uparrow+N_\downarrow)D_\downarrow})^{-1}$ is 
the spin averaged diffusion constant. Eq. \ref{diffusion} describes that the 
difference in $\mu$ decays over a length scale $\lambda_{sf}=\sqrt{D\tau_{sf}}$
, the spin-flip length.

To describe the $F/S$ system the role of the superconductor has to be 
incorporated. We assume that the interface 
resistance itself can be ignored, which is justified in metallic diffusive 
systems with transparent interfaces. The Andreev reflection can then be taken 
into account by the following boundary conditions at the F/S interface ($x=0$):
\begin{eqnarray}
\mu_{\uparrow}\arrowvert_{x=0} & = & -\mu_{\downarrow}\arrowvert_{x=0} 
\label{cond1} \\
j_{\uparrow}\arrowvert_{x=0} & = & j_{\downarrow}\arrowvert_{x=0}. 
\label{cond2}
\end{eqnarray}
Here the electrochemical potential of the superconductor S is set to zero. Eq. 
\ref{cond1} is a direct consequence of $AR$, where an excess of electrons with 
spin-up corresponds to an excess of holes and therefore a deficit of electrons 
with spin-down and vice versa.  Eq. \ref{cond2} arises due to the fact that the 
total Cooper pair spin in the superconductor is zero, so there can be no net 
spin current across the interface. Note that for Eqs.  \ref{cond1} and 
\ref{cond2} to be valid, no spin-flip processes are assumed to occur at the 
interface as well as in the superconductor.

Eqs. \ref{current} ,\ref{diffusion}, \ref{cond1} and \ref{cond2} now allow the 
calculation of the spatial dependence of the electrochemical potentials of both 
spin directions, which have the general forms:
\begin{equation}
\mu_{\uparrow}~=~A+Bx+\frac{C}{\sigma_{\uparrow}}e^{x/\lambda_{sf}^F}
+\frac{D}{\sigma_{\uparrow}}e^{-x/\lambda_{sf}^F}
\label{general+}
\end{equation}
\begin{equation}
\mu_{\downarrow}~=~A+Bx-\frac{C}{\sigma_{\downarrow}}e^{x/\lambda_{sf}^F}
-\frac{D}{\sigma_{\downarrow}}e^{-x/\lambda_{sf}^F} 
\label{general-}
\end{equation} where A,B,C and D are constants defined by the boundary 
conditions. For simplicity we first calculate the contact resistance at the 
F/S interface in a two terminal configuration, noted by $V_{2T}$  
in Fig. \ref{structure}(a), ignoring the presence of the second ferromagnetic 
electrode F2. In this configuration we find: 
\begin{equation}
\mu_{\uparrow}\arrowvert_{x=0}=-\mu_{\downarrow}\arrowvert_{x=0}=\frac
{\alpha_F\lambda_{sf}^FeI}{\sigma_F(1-\alpha_F^2)A} 
\label{mus}
\end{equation} where 
$\alpha_F=(\sigma_\uparrow - \sigma_\downarrow)/(\sigma_\uparrow + 
\sigma_\downarrow)$ is the spin polarization of the current 
in the bulk ferromagnet and $\lambda_{sf}^F$, $\sigma_F=\sigma_\uparrow + 
\sigma_\downarrow$, $A$ are the spin-flip length, the conductivity and the 
cross-sectional area of the ferromagnetic strip, respectively. Note that at 
the interface the electrochemical potentials are finite, despite the presence 
of the superconductor. This is illustrated in the left part of 
Fig. \ref{chemFS}, where the spin-up and 
spin-down electrochemical potentials are plotted as a function of x in units of 
$\lambda_{sf}^F$. Defining a contact resistance as $R_{FS}=
\Delta\mu/eI$ at the F/S interface yields\cite{fal1}:
\begin{equation}
R_{FS} ~=~ \frac{\alpha_F^2\lambda_{sf}^F}{\sigma_F(1-\alpha_F^2)A}.
\label{fs}
\end{equation} 

Note that this is exactly half the resistance which would be measured in a two 
terminal geometry of one ferromagnetic electrode directly coupled to another 
ferromagnetic electrode with anti-parallel magnetization. One may 
therefore consider the F/S interface as an 'ideal' domain wall (which does not 
change the spin direction), the superconductor acting as a magnetization 
mirror.

The presence of the contact resistance at a F/S boundary clearly brings out the 
difference between a superconductor and a normal conductor with infinite 
conductivity. In the latter case  the boundary condition Eq. \ref{cond1} at the 
interface is replaced by $\mu_\uparrow=\mu_\downarrow=0$, and no contact 
resistance would be generated\cite{joh1,son1}. An interesting feature to be 
noticed from Fig. \ref{chemFS} is that the electrochemical potential of the 
minority spin at the interface is \emph{negative}.  

The second observation to be made here is that the excess charge density 
$n_c \sim \mu_\uparrow + \mu_\downarrow$ is zero, whereas the spin density $n_s 
\sim \mu_\uparrow - \mu_\downarrow$ has a maximum close to the interface. This 
is a direct consequence of the $AR$ process, where a net spin current is not 
allowed to enter the superconductor. Continuity of the spin currents at the F/S 
interface results in a spin accumulation in the ferromagnet, being build up 
over a distance of the spin-flip length $\lambda_{sf}^F$. 

The contact resistance is small ($R_{FS}\approx 20~m\Omega$ for a nickel 
strip with a thickness of 20 nm, a width of 100 nm, a resistivity of 
$\sim 80\times 10^{-9}~\Omega m$, a spin polarization $\alpha_F\approx 0.2$ and 
a spin-flip length of 20 nm\cite{soulen1,upadhyay1,ted1,onbek2,ferrosf}) 
compared to the total resistance of the ferromagnetic strip F1. 

To identify the small contact resistance it is necessary to use a 
multi-terminal geometry. The four terminal resistance is measured by sending a 
current through terminals 1 and 3, and measuring the voltage between terminals 
2 and 4, as illustrated by $V_{4T}$ in Fig. \ref{structure}(a). We 
assume that all current flows into the 
superconductor at $x=0$, which is reasonable to assume when the thickness $d_F$ 
of the ferromagnetic strip is small compared to the width $W$ of the 
superconductor (cf. Fig. \ref{structure}(b)). The width $W$ of the 
superconductor 
is assumed to be smaller than the spin-flip length of the ferromagnetic strip, 
$W<\lambda_{sf}^F$. Now the second ferromagnetic electrode (F2) has to be 
included in the calculation. This is done by requiring Eqs. \ref{cond1} and 
\ref{cond2} to 
include the spin currents of both ferromagnetic electrodes and requiring their 
spin-up and down-spin electrochemical potentials to be continuous. For the 
resistance in the four terminal geometry of Fig. \ref{structure} the 
calculation yields:
\begin{equation}
R_{FS'} ~=~\pm\frac{1}{2}~\frac{\alpha_F^2\lambda_{sf}^F}{\sigma_F(1-
\alpha_F^2)A} 
\label{fs2}
\end{equation} where the  sign refers to the parallel (+) or anti-parallel (-) 
relative orientation of the magnetization of the two ferromagnetic electrodes. 
In the case of anti-parallel arrangement one therefore has the rather unique 
situation that the voltage measured can be outside the range of source and 
drain contacts\cite{foot2}.

The above holds as long as the spin-flip length $\lambda_{sf}^F$ exceeds the 
width $W$ of the superconductor. The complication of the above experiment 
would be that it requires the width of the superconductor to be shorter than 
the spin-flip length in the ferromagnet, which is expected to be around 20 
nm\cite{ferrosf}. To remedy these complications, we consider an alternative 
geometry. 
 
The geometry (F/N/S) of Fig. \ref{structure2} consists of two superconducting 
strips S, which are coupled by a thin layer of normal metal N, which has a 
larger spin-flip length ($\lambda_{sf}^N$) than the spin-flip length of the 
ferromagnet ($\lambda_{sf}^F$)\cite{joh1}. On top of the normal metal two 
ferromagnetic strips F1 and F2 are placed. Current is injected by F1 through 
the normal metal, into the superconductor, whereas the voltage is detected by 
F2. 

In the absence of a spin polarized current $I$, the measured resistance 
$R=V/I$, will decay exponentially with $R_0exp(-CL/d_N)$, where 
$R_0\approx\rho_Nd_N/A_C$ is the resistance of the normal metal between the 
superconductor and the current injector F1. Here $\rho_N$ is the resistivity of 
the normal metal, $A_C$ the contact area between F1 and S, $d_N$ the thickness 
of the normal metal, $C$ a constant of order unity and $L$ the distance 
between the two ferromagnetic strips. This resistance will therefore vanish 
in the regime $L \gg d_N$. However, in the presence of a spin-polarized 
current $I$ a spin density is created at the current injector F1, stretching 
out towards the voltage probe F2.  

To calculate the signal at F2 we have to include the normal region. First, we 
assume that the superconductor in the region S$^{'}$ in Fig. \ref{structure2} 
is absent. We take the non-equilibrium spin density to be uniform in the normal 
metal in the region under F1, which is allowed as the thickness of the normal 
metal is small compared to the spin-flip length ($\lambda_{sf}^N$) in the 
normal metal, $~d_N\ll\lambda_{sf}^N~$. The electrochemical potentials in the 
normal region between the two ferromagnetic strips are described by solutions 
of Eq. \ref{general+} and \ref{general-}, with the constants $A=B=0$. We then 
calculate the resistance in the relevant limit that the distance $L$ does not 
exceed the spin-flip length of the normal region, $L\lesssim\lambda_{sf}^N$. 
The expression for the resistance in this limit is given by:  
\begin{equation}
R_{FNS}=\pm\frac{\alpha_F^2\lambda_{sf}^F}{2\sigma_FA(1-\alpha_F^2)+\frac
{L\sigma_F^2A}{\sigma_{N}\lambda_{sf}^F}(1+\alpha_F)^2(1-\alpha_F)^2}
\label{FNS}
\end{equation}where $\sigma_N$ is the conductivity of the normal metal and $L$  
is the distance between the two ferromagnetic electrodes. When 
$L>\lambda_{sf}^N$ the signal will decay exponentially. 

Eq. \ref{FNS} and Fig. \ref{chemFNS} show that, even though no charge 
current flows in the N layer, nevertheless a signal is generated at the 
ferromagnetic electrode F2. In addition, Eq. \ref{FNS} shows that the 
signal changes sign when the polarization of F2 is reversed. A 
reduction of the thickness of the N film will reduce the 
signal. This is a consequence of the fact that although no charge current 
flows, the spin-up and spin-down currents are non-zero, and their magnitude 
(and the associated voltage) depends on the resistance of the N layer. 

The above analysis is based on classical assumptions, where the superconducting 
proximity effect has been ignored in the normal metal. However, it is known 
that a superconductor modifies the electronic states in the N layer\cite
{belzig1,gueron1}, which would be the case when a superconductor is present in 
the region S$^{'}$ (cf. Fig. \ref{structure2}).

In this situation Eq. \ref{FNS} would still hold, for the electrochemical 
potentials in the normal metal satisfy the boundary condition of Eq.  
\ref{cond1}. When the thickness $d_N$ of the normal layer is of the order of 
the superconducting coherence length $\xi$, a gap $\Delta_N$ will be 
developed in the normal metal. This will prohibit the opposite spin currents 
in the normal metal to flow, and therefore  no signal will be detected at 
the ferromagnetic electrode F2. One could control and eliminate the 
induced gap $\Delta_N$ by applying a magnetic field parallel to the ferromagnetic 
electrodes. 

To conclude, we have shown that the spin reversal associated with Andreev 
reflection in a diffusive ferromagnet-superconductor junction, leads to a 
spin contact resistance. The contact resistance is due to an excess spin 
density, which exists close to the F/S interface, on a length scale of the 
spin-flip length in the ferromagnet. In a multi-terminal geometry the  
contact resistance can have a positive and negative sign, depending on the 
relative orientation of the ferromagnetic electrodes.

The authors wish to thank the Stichting Fundamenteel Onderzoek der Materie and 
the EU ESPRIT project no 23307 SPIDER for financial support.


%
\begin{figure}[tbp] 
\caption{(a) Top view of a cross type F/S geometry. S is the 
superconducting strip on top of two ferromagnetic strips F1 and F2. The 
magnetization of F2 can be parallel or anti-parallel to the magnetization of 
F1. The x-axis is taken along the ferromagnetic strips, where from $x=0$ 
to $x=W$ the superconducting strip covers the ferromagnetic strips. (b) 
Side view.}
\label{structure} 
\end{figure}
\begin{figure}[tbp] 
\caption{Electrochemical potential in the ferromagnetic strip 
of Fig. \ref{structure} as a function of distance along the x-axis in units 
of the spin-flip length $\lambda_{sf}^F$. The potential of the superconductor 
at $x=0$ is set to zero. The solid curves at $x>0$ yield the chemical 
potentials for the two spin directions when the ferromagnetic electrode F2 is 
magnetized parallel to the magnetization of F1. The dotted curves yield the 
electrochemical potentials for anti-parallel magnetization.}
\label{chemFS}
\end{figure} 
\begin{figure}[tbp]
\caption{(a) Top view of a F/N/S geometry. N is a normal metal strip coupling 
to the two superconducting strips S. In the region S$^{'}$ a superconductor 
maybe  present (see text). On top of the normal metal two ferromagnetic strips 
F1 and F2 are placed. (b) Side view, terminals 3 and 1 are used for current 
injection and extraction, whereas terminals 2 and 4 measure the voltage. 
M refers to the magnetization of the ferromagnetic electrodes F1 and F2. 
$L$ is the distance between the two ferromagnetic electrodes and $d_N$ is 
the thickness of the normal metal.}
\label{structure2}
\end{figure}
\begin{figure}[tbp] 
\caption{Electrochemical potential versus distance. The 
coordinate $x=0$ defines the position of the ferromagnetic electrode F1. 
The coordinate $x=L=2\lambda_{sf}^F$ defines the position of the ferromagnetic
electrode F2. The solid curves for $x>L$ yield the chemical potentials for 
the two spins when the ferromagnetic electrode F2 is magnetized parallel to the 
magnetization of F1. The dotted curves yield the chemical potentials for 
anti-parallel magnetization.}
\label{chemFNS} 
\end{figure}


\begin{references}
\bibitem{foot0}In our analysis of the F/S interface we only take into account 
the electron transport below $\Delta$. This distinguishes our work from the 
studies of spin injection in superconductors, which can only occur for energies 
above $\Delta$.
\bibitem{onbek2} The value of $\alpha_F$ can only be estimated in relation 
with the values taken from the tunnel junction and point contact 
experiments\cite{soulen1,upadhyay1,ted1}. Since in 
our description the spin polarization of the current is not solely determined 
by the density of states at the Fermi level, but also includes the different 
spin diffusion constants.
\bibitem{foot2} This is made possible by the Andreev reflection provided by 
the superconductor. In absence of the  superconductor, one would 
always measure an electrochemical potential between the source and drain 
contacts.
\end{references}
\end{document}